\documentclass[aps,prd,preprint,superscriptaddress,nofootinbib]{revtex4-2}
\usepackage{amsmath,amssymb,mathtools}
\usepackage{booktabs}
\usepackage{array}
\usepackage{hyperref}
\usepackage{microtype}
\hypersetup{hidelinks}
\newcommand{\appref}[1]{\hyperref[#1]{Appendix~\ref*{#1}}}
\allowdisplaybreaks

\makeatletter
\def\frontmatter@abstract@produce{%
  \par
  \addvspace{\frontmatter@preabstractspace}%
  \begingroup
    \dimen@\baselineskip
    \setbox\z@\vtop{\unvcopy\absbox}%
    \advance\dimen@-\ht\z@\advance\dimen@-\prevdepth
    \@ifdim{\dimen@>\z@}{\vskip\dimen@}{}%
  \endgroup
  \begingroup
    \prep@absbox
    \unvbox\absbox
    \post@absbox
  \endgroup
  \@ifx{\@empty\mini@notes}{}{\mini@notes\par}%
  \addvspace\frontmatter@postabstractspace
}%
\makeatother

\newcommand{\dd}{\mathrm{d}}
\newcommand{\ii}{\mathrm{i}}
\newcommand{\ee}{\mathrm{e}}
\newcommand{\Rloc}{\mathcal{R}_{\mathrm{loc}}}
\newcommand{\RAB}{\mathcal{R}_{AB}}

\newcommand{\sinc}{\operatorname{sinc}}
\newcommand{\asinh}{\operatorname{arsinh}}
\newcommand{\csch}{\operatorname{csch}}
\newcommand{\dS}{\mathrm{dS}}
\newcommand{\acc}{\mathrm{acc}}
\newcommand{\thm}{\mathrm{th}}
\newcommand{\Mink}{\mathrm{M}}

\begin{document}

\title{Comparison of Two-Atom Cross Spectra in de Sitter Spacetime, Uniformly Accelerated Minkowski Vacuum, and a Thermal Bath}
\author{Zhiming Huang}
\email[Corresponding author: ]{465609785@qq.com}
\affiliation{School of Economics and Management, Wuyi University, Jiangmen 529020, China}

\begin{abstract}
We study the weak coupling of two identical two-level atoms to a four-dimensional massless conformally coupled scalar field and compare three settings with the same local temperature but different physical origins: comoving atoms in the Bunch--Davies vacuum of de Sitter spacetime, transversely separated uniformly accelerated atoms in the Minkowski vacuum, and static atoms in a Minkowski thermal bath. We first derive the two-atom cross spectra in the three settings and then obtain the single-atom local spectra uniformly from the zero-separation limit. When $a=H$ and $\beta=2\pi/H$, the complete local Wightman kernels and local spectra are identical in all three cases. At finite separation, a genuine thermal bath produces a $\sinc(\Omega L)$ spatial factor, whereas de Sitter spacetime and the transversely accelerated vacuum produce a hyperbolic geometric factor. In the long-distance limit, the cross correlations in de Sitter spacetime and the transversely accelerated vacuum decay as $L^{-2}$, whereas those in the Minkowski thermal bath decay only as $L^{-1}$. The de Sitter and transversely accelerated vacuum cross spectra coincide when $a=H$ and the separations are instantaneously matched. For a fixed experimental setup, the physical separation between the two comoving atoms in de Sitter space evolves with cosmic expansion, whereas the pulled-back cross correlations in the uniformly accelerated and thermal Minkowski configurations are stationary.
\end{abstract}

\maketitle

\section{Introduction}

A freely falling detector in de Sitter spacetime measures the Gibbons--Hawking temperature in the Bunch--Davies vacuum
\begin{equation}
T_{\mathrm{GH}}=\frac{H}{2\pi},
\end{equation}
whereas a uniformly accelerated detector in the Minkowski vacuum measures the Unruh temperature
\begin{equation}
T_U=\frac{a}{2\pi}.
\end{equation}
When $a=H$, the two local thermal responses can coincide; if the inverse temperature of the Minkowski thermal bath is further chosen as $\beta=2\pi/H$, the third setting has the same local temperature as well\cite{GibbonsHawking1977,Unruh1976,DeserLevin1997,Crispino2008,TianJing2014}. Therefore, the local response along a single worldline cannot by itself identify whether the thermality originates from spacetime curvature, noninertial motion, or a genuine thermal environment.

A two-atom system also contains the cross Wightman function between distinct spatial points
\begin{equation}
G^+_{AB}(\tau,\tau')
=\langle\phi[x_A(\tau)]\phi[x_B(\tau')]\rangle,
\label{eq:intro-cross}
\end{equation}
which depends on both the temporal and spatial relations between the two worldlines. Previous studies of collective transitions, entanglement, and interatomic interactions have shown that such nonlocal correlations directly affect two-atom dynamics in de Sitter spacetime, accelerated vacuum, and thermal baths\cite{HuYu2013,Salton2015,HuangTian2017,ZhouYu2020,ZhouChengYu2023,ZhouZhou2024,Shaukat2026}. Here we focus on a more specific question: after the local temperatures are matched exactly, do the cross two-point spectra in the three settings remain different, and can these differences be compared analytically?

We compare three configurations: two comoving geodesics in the spatially flat patch of de Sitter spacetime; two worldlines in the Minkowski vacuum with the same proper acceleration and a fixed transverse separation; and two static atoms in a Minkowski thermal bath. We impose throughout
\begin{equation}
a=H,
\qquad
\beta=\frac{2\pi}{H}.
\label{eq:intro-matching}
\end{equation}
The main results are as follows. In the zero-separation limit, the three cross spectra yield exactly the same local Wightman kernel and local spectrum. At finite separation, the thermal bath carries the ordinary $\sinc$ spatial factor, whereas the two vacuum settings share the same hyperbolic geometric factor. The de Sitter and transversely accelerated cases exhibit an exact cross-spectrum degeneracy when their separations are instantaneously matched, but the physical separation of the comoving de Sitter atoms changes with time, so this degeneracy cannot persist for fixed experimental configurations.

We use $\hbar=c=k_B=1$ throughout and adopt the Fourier convention
\begin{equation}
\mathcal R(\Omega)
=\int_{-\infty}^{\infty}\dd s\,
\ee^{-\ii\Omega s}G^+(s),
\qquad \Omega>0.
\label{eq:fourier-convention}
\end{equation}
If the two-point kernel pulled back to the worldline depends only on the relative time $s$, Eq.~\eqref{eq:fourier-convention} defines a stationary Fourier spectrum. The cross kernel of the comoving de Sitter atoms also depends on the midpoint time $T$; hence $\mathcal R_{AB}(T,\Omega)$ in this work should be understood as the two-point spectrum obtained by Fourier transforming the relative time at fixed $T$, rather than as an infinite-time transition rate for a general nonstationary process.

\section{Two-atom model and unified cross-spectrum framework}

Consider two identical two-level atoms $A$ and $B$ with energy gap $\Omega>0$. Their free Hamiltonian and monopole operators in the interaction picture are, respectively,
\begin{equation}
H_S
=\frac{\Omega}{2}\sigma_3^{(A)}
+\frac{\Omega}{2}\sigma_3^{(B)},
\label{eq:atom-H}
\end{equation}
\begin{equation}
m_\mu(\tau)
=\sigma_\mu^+\ee^{\ii\Omega\tau}
+\sigma_\mu^-\ee^{-\ii\Omega\tau},
\qquad \mu=A,B.
\label{eq:monopole}
\end{equation}
The atoms are weakly and continuously coupled to the field through the interaction Hamiltonian
\begin{equation}
H_I(\tau)
=\lambda\sum_{\mu=A,B}
m_\mu(\tau)\phi[x_\mu(\tau)],
\label{eq:interaction}
\end{equation}
where $\lambda$ is a small coupling constant and $\phi$ is a four-dimensional massless conformally coupled real scalar field\cite{DeWitt1979,BirrellDavies1982}.

Define the Wightman matrix
\begin{equation}
G^+_{\mu\nu}(\tau,\tau')
=\langle\phi[x_\mu(\tau)]\phi[x_\nu(\tau')]\rangle,
\qquad \mu,\nu=A,B.
\label{eq:wightman-matrix}
\end{equation}
The diagonal entries describe local correlations, whereas the off-diagonal entries describe cross correlations between the two atoms. For the symmetric configurations considered here, $G^+_{AA}=G^+_{BB}\equiv G^+_{\mathrm{loc}}$. In stationary cases, we define the spectral matrix
\begin{equation}
\mathcal R_{\mu\nu}(\Omega)
=\int_{-\infty}^{\infty}\dd s\,
\ee^{-\ii\Omega s}G^+_{\mu\nu}(s),
\qquad s=\tau-\tau'.
\label{eq:spectral-matrix}
\end{equation}
For the comoving de Sitter atoms, introduce
\begin{equation}
T=\frac{\tau+\tau'}{2},
\qquad s=\tau-\tau',
\label{eq:midtime}
\end{equation}
and define
\begin{equation}
\mathcal R_{AB}(T,\Omega)
=\int_{-\infty}^{\infty}\dd s\,
\ee^{-\ii\Omega s}G^+_{AB}(T,s).
\label{eq:wigner-cross}
\end{equation}
Here $T$ is retained as an external parameter and the Fourier transform acts only on the relative time $s$. Thus $\mathcal R_{AB}(T,\Omega)$ characterizes the frequency structure at a given midpoint time and may also be viewed as a local cross spectrum in the Wigner--Fourier representation. For a general nonstationary process, a finite-time response requires integrating the switching function together with the full $G^+_{AB}(T,s)$ and cannot be obtained from this fixed-$T$ Fourier transform alone.

The three sets of worldlines are chosen as follows. In spatially flat de Sitter coordinates,
\begin{equation}
\dd s^2=\dd t^2-\ee^{2Ht}\dd\vec x^{\,2}
=a^2(\eta)(\dd\eta^2-\dd\vec x^{\,2}),
\qquad
a(\eta)=-\frac1{H\eta},
\label{eq:dsmetric}
\end{equation}
The two atoms are fixed at $\vec x_A=(0,0,0)$ and $\vec x_B=(L_c,0,0)$. Along the comoving worldlines, $\tau=t$. Here $L_c$ is the time-independent comoving coordinate separation; the corresponding equal-time physical separation is determined by the de Sitter scale factor, as derived explicitly in \appref{app:ds-kernel}. For uniform acceleration in Minkowski spacetime, we choose
\begin{align}
t_A(\tau)&=a^{-1}\sinh(a\tau),
&x_A(\tau)&=a^{-1}\cosh(a\tau),
&y_A&=0,\notag\\
t_B(\tau)&=a^{-1}\sinh(a\tau),
&x_B(\tau)&=a^{-1}\cosh(a\tau),
&y_B&=L,
\label{eq:acc-trajectories}
\end{align}
so that the two atoms have the same proper acceleration and maintain a fixed transverse separation $L$. In the thermal Minkowski case, the two atoms are static and separated by $L$. Since proper time on a static worldline coincides with Minkowski coordinate time, we may take $\tau=t$; the field is in a Gibbs state of inverse temperature $\beta$.

\section{Cross spectra, local degeneracy, and temporal evolution in the three settings}

In the Bunch--Davies vacuum of de Sitter spacetime, substituting the two comoving worldlines into the Wightman function of the conformal scalar field gives
\begin{equation}
G^{+}_{AB,\dS}(T,s)
=-\frac{H^2}{16\pi^2}
\frac{1}{
\sinh^2\!\left[\dfrac{H(s-\ii0^+)}{2}\right]
-\left[\dfrac{Hr(T)}{2}\right]^2
}.
\label{eq:Gds-cross}
\end{equation}
where $r(T)$ denotes the equal-time physical separation of the two comoving worldlines at the midpoint time $T=(\tau+\tau')/2$.
At fixed $T$, Fourier transformation with respect to $s$ gives, by contour integration,
\begin{equation}
\mathcal R_{AB}^{\dS}(T,\Omega)
=
\frac{
\sin\!\left[
\dfrac{2\Omega}{H}
\asinh\!\left(\dfrac{Hr(T)}2\right)
\right]
}{2\pi r(T)\sqrt{1+[Hr(T)/2]^2}}
\frac1{\ee^{2\pi\Omega/H}-1}.
\label{eq:Rds-direct}
\end{equation}
The cross kernel and residue calculation are given in \appref{app:ds-kernel} and \appref{app:ds-contour}, respectively.

For two transversely separated uniformly accelerated atoms in the Minkowski vacuum, the individual coordinate differences contain the midpoint time $T$, but $T$ cancels completely when they are combined into the Minkowski invariant. The cross kernel therefore depends only on the relative time $s$, and the configuration is stationary. Thus
\begin{equation}
G^+_{AB,\acc}(s)
=-\frac{a^2}{16\pi^2}
\frac{1}{
\sinh^2\!\left[\dfrac{a(s-\ii0^+)}{2}\right]
-\left(\dfrac{aL}{2}\right)^2
}.
\label{eq:Gacc-cross}
\end{equation}
Its stationary cross spectrum is
\begin{equation}
\mathcal R_{AB}^{\acc}(\Omega)
=
\frac{
\sin\!\left[
\dfrac{2\Omega}{a}
\asinh\!\left(\dfrac{aL}{2}\right)
\right]
}{2\pi L\sqrt{1+(aL/2)^2}}
\frac1{\ee^{2\pi\Omega/a}-1}.
\label{eq:Racc-direct}
\end{equation}
A detailed derivation is given in \appref{app:acc-spectrum}.

For a Minkowski thermal bath of inverse temperature $\beta$, the mode expansion and angular integration give
\begin{equation}
\mathcal R_{AB}^{\thm}(\Omega,L)
=\frac{\Omega}{2\pi}
\frac1{\ee^{\beta\Omega}-1}
\frac{\sin(\Omega L)}{\Omega L}.
\label{eq:Rth-cross}
\end{equation}
A detailed derivation is given in \appref{app:thermal-spectrum}.

The single-atom local spectra can now be obtained uniformly from the zero-separation limits of the three cross spectra. Using $\asinh x=x+O(x^3)$, $\sin x=x+O(x^3)$, and $\lim_{x\to0}\sin x/x=1$, we obtain
\begin{align}
\lim_{r\to0}\mathcal R_{AB}^{\dS}(T,\Omega)
&=\frac{\Omega}{2\pi}
\frac1{\ee^{2\pi\Omega/H}-1},\notag\\
\lim_{L\to0}\mathcal R_{AB}^{\acc}(\Omega)
&=\frac{\Omega}{2\pi}
\frac1{\ee^{2\pi\Omega/a}-1},\notag\\
\lim_{L\to0}\mathcal R_{AB}^{\thm}(\Omega,L)
&=\frac{\Omega}{2\pi}
\frac1{\ee^{\beta\Omega}-1}.
\label{eq:three-local-limits}
\end{align}
After setting $a=H$ and $\beta=2\pi/H$, all three reduce to the same local spectrum
\begin{equation}
\mathcal R_{\mathrm{loc}}^{\dS}
=\mathcal R_{\mathrm{loc}}^{\acc}
=\mathcal R_{\mathrm{loc}}^{\thm}
\equiv\mathcal R_{\mathrm{loc}}(\Omega)
=\frac{\Omega}{2\pi}
\frac1{\ee^{2\pi\Omega/H}-1}.
\label{eq:triple-local}
\end{equation}
The degeneracy in fact already occurs at the level of the two-point kernels:
\begin{equation}
G^+_{\mathrm{loc},\dS}(s)
=G^+_{\mathrm{loc},\acc}(s)
=G^+_{\mathrm{loc},\thm}(s)
=-\frac{H^2}{16\pi^2}
\csch^2\!\left[\frac{H(s-\ii0^+)}2\right].
\label{eq:triple-local-kernel}
\end{equation}
The image-sum derivation of the thermal local kernel is given in \appref{app:thermal-spectrum}. Thus the equality of the local spectra is not an accidental coincidence of three independent Planck factors, but follows from the pointwise identity of the local Wightman kernels.

To separate the local response from the nonlocal modification induced by finite separation, introduce the spatial modulation factors
\begin{equation}
f_H(\Omega,r)
=
\frac{
\sin\!\left[
\dfrac{2\Omega}{H}
\asinh\!\left(\dfrac{Hr}{2}\right)
\right]
}{\Omega r\sqrt{1+(Hr/2)^2}},
\label{eq:fH-main}
\end{equation}
and
\begin{equation}
f_a(\Omega,L)
=
\frac{
\sin\!\left[
\dfrac{2\Omega}{a}
\asinh\!\left(\dfrac{aL}{2}\right)
\right]
}{\Omega L\sqrt{1+(aL/2)^2}}.
\label{eq:fa-main}
\end{equation}
After temperature matching,
\begin{align}
\mathcal R_{AB}^{\dS}(T,\Omega)
&=\mathcal R_{\mathrm{loc}}(\Omega)f_H[\Omega,r(T)],
\label{eq:Rds-factor}\\
\mathcal R_{AB}^{\acc}(\Omega)
&=\mathcal R_{\mathrm{loc}}(\Omega)f_H(\Omega,L),\notag\\
\mathcal R_{AB}^{\thm}(\Omega,L)
&=\mathcal R_{\mathrm{loc}}(\Omega)\sinc(\Omega L),
\label{eq:factorized-three}
\end{align}
where $f_a=f_H$ when $a=H$.

Equations~\eqref{eq:Rds-factor}--\eqref{eq:factorized-three} show that the three settings have the same local spectrum, while their finite-separation differences are determined entirely by the corresponding spatial modulation factors.

For $HL\ll1$, the leading correction of the geometric factor relative to the thermal result is
\begin{equation}
f_H(\Omega,L)
=\sinc(\Omega L)
-(HL)^2\left[
\frac18\sinc(\Omega L)
+\frac1{24}\cos(\Omega L)
\right]
+O((HL)^4).
\label{eq:small-HL}
\end{equation}
The derivation of this short-distance expansion and the corresponding long-distance asymptotics is given in \appref{app:spatial-asymptotics}. For $HL\gg1$, the spatial modulation factors in de Sitter spacetime and the transversely accelerated vacuum have an amplitude envelope proportional to $L^{-2}$, whereas the $\sinc$ factor in the Minkowski thermal bath decays as $L^{-1}$.

Equations~\eqref{eq:Gds-cross} and \eqref{eq:Gacc-cross} also imply a stronger relation. When
\begin{equation}
a=H,
\qquad
L=r(T),
\label{eq:instant-match}
\end{equation}
we have
\begin{equation}
G^+_{AB,\dS}(T,s)
=G^+_{AB,\acc}(s)
\big|_{a=H,\,L=r(T)},
\label{eq:kernel-flow-identity}
\end{equation}
and therefore
\begin{equation}
\mathcal R_{AB}^{\dS}(T,\Omega)
=\mathcal R_{AB}^{\acc}(\Omega)
\big|_{a=H,\,L=r(T)}.
\label{eq:spectral-flow-identity}
\end{equation}
Thus, when $a=H$ and $L=r(T)$, the comoving de Sitter atoms and the transversely accelerated atoms have the same cross kernel and cross spectrum at that instant.

Choose a reference time $T_0$ and set $L_0=r(T_0)$, and take the fixed separations in the other two configurations to be $L_0$ as well. From
\begin{equation}
r(T)=L_0\ee^{H(T-T_0)},
\label{eq:r-relative}
\end{equation}
which gives
\begin{align}
f_H[\Omega,r(T)]
&=f_H\!\left(\Omega,L_0\ee^{H(T-T_0)}\right),
\label{eq:Cds}\\
f_H(\Omega,L_0)
&=f_H(\Omega,L_0),
\label{eq:Cacc}\\
\sinc(\Omega L_0)
&=\sinc(\Omega L_0).
\label{eq:Cth}
\end{align}
At $T=T_0$, $r(T_0)=L_0$, so the spatial modulation factors of the two vacuum settings are identical. Thereafter, the transversely accelerated and thermal configurations remain unchanged, whereas the spatial modulation factor of the comoving de Sitter configuration changes as $r(T)$ grows, approaching $1$ at early times and $0$ at late times.

\section{Conclusion}

We have compared comoving atoms in de Sitter spacetime, transversely separated uniformly accelerated atoms in Minkowski spacetime, and static atoms in a Minkowski thermal bath under matched local temperatures. The main results can be summarized in three points. First, when $a=H$ and $\beta=2\pi/H$, the local Wightman kernels and local spectra are exactly identical in the three settings. Second, at finite separation this threefold degeneracy is lifted: the Minkowski thermal bath yields a $\sinc(\Omega L)$ spatial factor, whereas de Sitter spacetime and the transversely accelerated vacuum yield the same hyperbolic geometric factor. In particular, for $HL\gg1$, the spatial modulation factors in de Sitter spacetime and the transversely accelerated vacuum have an $L^{-2}$ amplitude envelope, whereas the thermal $\sinc$ factor decays as $L^{-1}$. Third, when $a=H$ and the instantaneous physical separation satisfies $L=r(T)$, the de Sitter and transversely accelerated vacuum cross kernels and cross spectra are identical; however, for fixed configurations the physical separation of the comoving de Sitter atoms changes with the expansion, so this equivalence can hold only at a single matching instant.

These results apply to a four-dimensional massless conformally coupled scalar field, pointlike two-level detectors, and the worldline geometries considered here. The static de Sitter comparison in \appref{app:static-ds} further shows that the cross spectrum depends not only on the field state but also on the detector worldlines and spatial configuration.

\appendix

\section{de Sitter cross kernel for two comoving atoms}\label{app:ds-kernel}

The four-dimensional spatially flat de Sitter metric is
\begin{equation}
\dd s^2
=\dd t^2-\ee^{2Ht}\dd\vec x^{\,2}.
\label{app:ds-cosmic}
\end{equation}
Introduce conformal time $\eta$, satisfying
\begin{equation}
\dd\eta=\ee^{-Ht}\dd t.
\end{equation}
Integrating and choosing the expanding branch with $\eta<0$ gives
\begin{equation}
\eta=-\frac{\ee^{-Ht}}{H},
\qquad
a(\eta)=\ee^{Ht}=-\frac1{H\eta}.
\label{app:eta}
\end{equation}
Therefore
\begin{equation}
\dd s^2=a^2(\eta)
\left(\dd\eta^2-\dd\vec x^{\,2}\right).
\label{app:conformalmetric}
\end{equation}

We consider a four-dimensional massless conformally coupled scalar field ($\xi=1/6$). Since the spatially flat de Sitter metric can be written as $g_{\mu\nu}=a^2(\eta)\eta^{\rm M}_{\mu\nu}$, where $\eta^{\rm M}_{\mu\nu}$ is the Minkowski metric, introduce the rescaled field
\begin{equation}
\widetilde\phi(\eta,\vec x)
\equiv a(\eta)\phi_{\dS}(\eta,\vec x)
\end{equation}
for which the massless conformally coupled Klein--Gordon equation reduces to the massless field equation in Minkowski spacetime
\begin{equation}
(\eta_{\rm M})^{\mu\nu}\partial_\mu\partial_\nu\widetilde\phi=0.
\end{equation}
Thus $\widetilde\phi$ can be expanded in the usual positive-frequency Minkowski modes; under this conformal mapping it can be identified with the corresponding Minkowski field $\phi_{\Mink}$, so that
\begin{equation}
\phi_{\dS}(\eta,\vec x)
=\frac1{a(\eta)}\phi_{\Mink}(\eta,\vec x).
\label{app:field-conformal}
\end{equation}
For a four-dimensional massless conformally coupled scalar field, the Bunch--Davies vacuum coincides with the conformal vacuum obtained by mapping these positive-frequency Minkowski modes as above\cite{BirrellDavies1982}. Since $\phi_{\dS}=a^{-1}\phi_{\Mink}$, the two-point function acquires one factor of $a^{-1}$ at each spacetime point, and hence the Wightman function satisfies
\begin{equation}
G^+_{\dS}(x,x')
=\frac1{a(\eta)a(\eta')}
G^+_{\Mink}(\eta,\vec x;\eta',\vec x').
\label{app:wightman-conformal}
\end{equation}
For the four-dimensional massless scalar field in Minkowski spacetime, we use the Wightman function
\begin{equation}
G^+_{\Mink}
=-\frac1{4\pi^2}
\frac1{(\eta-\eta'-\ii0^+)^2
-|\vec x-\vec x'|^2}.
\label{app:Mwightman}
\end{equation}
Hence
\begin{equation}
G^+_{\dS}(x,x')
=-\frac1{4\pi^2a(\eta)a(\eta')}
\frac1{(\eta-\eta'-\ii0^+)^2
-|\vec x-\vec x'|^2}.
\label{app:dswightman-general}
\end{equation}

The two comoving atoms are placed at
\begin{equation}
\vec x_A=(0,0,0),
\qquad
\vec x_B=(L_c,0,0).
\end{equation}
Here $L_c$ is the fixed comoving coordinate separation between the atoms. The physical distance should be evaluated on a spatial slice of equal cosmic time $t$, so we set $\dd t=0$. Equation~\eqref{app:ds-cosmic} then gives the proper spatial line element on that slice
\begin{equation}
\dd \ell^2=\ee^{2Ht}\dd\vec x^{\,2},
\qquad
\dd \ell=\ee^{Ht}|\dd\vec x|.
\label{app:spatial-line-element}
\end{equation}
Integrating along the $x$ direction connecting the two atoms, from $x=0$ to $x=L_c$, the physical distance at the same time $t$ is
\begin{align}
r(t)
&=\int_0^{L_c}\ee^{Ht}\dd x\notag\\
&=\ee^{Ht}L_c
=a(t)L_c.
\label{app:physical-distance-t}
\end{align}
Thus, although the comoving coordinate separation $L_c$ is fixed, the equal-time physical separation of the atoms grows with the scale factor $a(t)=\ee^{Ht}$.

Along a comoving worldline, $\dd\vec x=0$, and Eq.~\eqref{app:ds-cosmic} gives $\dd\tau=\dd t$; hence we may take $\tau=t$. Define
\begin{equation}
T=\frac{\tau+\tau'}2,
\qquad
s=\tau-\tau'.
\label{app:Ts}
\end{equation}
Equivalently,
\begin{equation}
\tau=T+\frac{s}{2},
\qquad
\tau'=T-\frac{s}{2}.
\label{app:inverse-Ts}
\end{equation}
Since $\tau=t$ along the comoving worldlines, the midpoint variable $T$ is also the arithmetic mean of the two cosmic times. Evaluating Eq.~\eqref{app:physical-distance-t} at $t=T$ gives the geometric distance parameter used below
\begin{equation}
r(T)=a(T)L_c=L_c\ee^{HT}.
\label{app:rT}
\end{equation}

Using Eq.~\eqref{app:eta}, the conformal times of the two events are
\begin{align}
\eta(\tau)
&=-\frac1H\ee^{-HT}\ee^{-Hs/2},\\
\eta(\tau')
&=-\frac1H\ee^{-HT}\ee^{Hs/2}.
\end{align}
The conformal-time difference is therefore
\begin{align}
\Delta\eta
&=\eta(\tau)-\eta(\tau')\\
&=\frac{\ee^{-HT}}H
\left(\ee^{Hs/2}-\ee^{-Hs/2}\right)\\
&=\frac{2\ee^{-HT}}H
\sinh\frac{Hs}{2}.
\label{app:deltaeta}
\end{align}
The product of the scale factors is
\begin{equation}
a(\tau)a(\tau')
=\ee^{H\tau}\ee^{H\tau'}
=\ee^{2HT}.
\label{app:aaproduct}
\end{equation}
Substituting into Eq.~\eqref{app:dswightman-general}, and temporarily suppressing the $i0^+$ prescription to make the algebra transparent, gives
\begin{align}
G^+_{AB,\dS}
&=-\frac{\ee^{-2HT}}{4\pi^2}
\frac1{
\dfrac{4\ee^{-2HT}}{H^2}
\sinh^2(Hs/2)-L_c^2}\\
&=-\frac{H^2}{16\pi^2}
\frac1{
\sinh^2(Hs/2)
-\dfrac{H^2L_c^2\ee^{2HT}}4}.
\end{align}
Using Eq.~\eqref{app:rT}, the spatial term in the denominator can be written as
\begin{equation}
\frac{H^2L_c^2\ee^{2HT}}{4}
=\left[\frac{Hr(T)}{2}\right]^2.
\label{app:space-term-rT}
\end{equation}
Restoring the Wightman boundary prescription then gives
\begin{equation}
G^{+}_{AB,\dS}(T,s)
=-\frac{H^2}{16\pi^2}
\frac{1}{
\sinh^2\!\left[\dfrac{H(s-\ii0^+)}2\right]
-\left[\dfrac{Hr(T)}2\right]^2}.
\end{equation}
This is Eq.~\eqref{eq:Gds-cross} in the main text.

When the two atoms coincide, $L_c=0$, equivalently $r(T)=0$, and the local kernel follows immediately:
\begin{equation}
G^+_{\mathrm{loc},\dS}(s)
=-\frac{H^2}{16\pi^2}
\frac1{
\sinh^2\!\left[\dfrac{H(s-\ii0^+)}2\right]}.
\label{app:Gds-local}
\end{equation}
It depends only on $s$, showing that the field fluctuations along a single comoving worldline are stationary; the nonstationarity arises only from the growth with $T$ of the physical separation between two distinct comoving points.

\section{Contour integral for the de Sitter cross spectrum}\label{app:ds-contour}

Start from
\begin{equation}
\RAB^{\dS}(T,\Omega)
=\int_{-\infty}^{\infty}\dd s\,
\ee^{-\ii\Omega s}G^+_{AB,\dS}(T,s)
\end{equation}
At fixed $T$, $r=r(T)$ is a constant parameter in the $s$ integration. Introduce the dimensionless variables
\begin{equation}
u=\frac{Hs}{2},
\qquad
\nu_0=\frac{2\Omega}{H},
\qquad
\rho=\frac{Hr}{2},
\qquad
\alpha=\asinh\rho.
\label{app:dimless}
\end{equation}
Since $\dd s=2\dd u/H$,
\begin{equation}
\RAB^{\dS}
=-\frac{H}{8\pi^2}
\int_{-\infty}^{\infty}\dd u\,
\frac{\ee^{-\ii\nu_0u}}
{\sinh^2(u-\ii0^+)-\rho^2}.
\label{app:udS-int}
\end{equation}

Ignoring the infinitesimal displacement for the moment, the poles are determined by
\begin{equation}
\sinh^2u=\rho^2
\end{equation}
i.e.,
\begin{equation}
\sinh u=\pm\rho.
\end{equation}
Because $\sinh(u+\ii\pi)=-\sinh u$, the complete set of poles can be written uniformly as
\begin{equation}
u_{n,+}=\alpha+\ii\pi n,
\qquad
u_{n,-}=-\alpha+\ii\pi n,
\qquad n\in\mathbb Z.
\label{app:poles}
\end{equation}
The original Wightman prescription acts on the time difference and shifts both light-cone poles on the real axis above the integration contour. For $\Omega>0$, let $u=x+\ii y$; then
\begin{equation}
|\ee^{-\ii\nu_0u}|=\ee^{\nu_0y}.
\end{equation}
Hence the exponential decays in the lower half-plane $y<0$, so the contour should be closed downward. The lower half-plane contains
\begin{equation}
u_{m,+}^{\downarrow}=\alpha-\ii\pi m,
\qquad
u_{m,-}^{\downarrow}=-\alpha-\ii\pi m,
\qquad m=1,2,\ldots
\end{equation}
which are the two sequences of poles.

Let
\begin{equation}
D(u)=\sinh^2u-\rho^2.
\end{equation}
Its derivative is
\begin{equation}
D'(u)=2\sinh u\cosh u.
\end{equation}
At $u=\alpha$,
\begin{equation}
\sinh\alpha=\rho,
\qquad
\cosh\alpha=\sqrt{1+\rho^2},
\end{equation}
so
\begin{equation}
D'(\alpha)=2\rho\sqrt{1+\rho^2}.
\end{equation}
At $u=-\alpha$,
\begin{equation}
D'(-\alpha)=-2\rho\sqrt{1+\rho^2}.
\end{equation}
Using $\sinh(u-\ii\pi m)=(-1)^m\sinh u$ and $\cosh(u-\ii\pi m)=(-1)^m\cosh u$, their product is unchanged, so the derivatives at each level $m$ retain the same respective signs.

Let
\begin{equation}
F(u)=\frac{\ee^{-\ii\nu_0u}}{D(u)}.
\end{equation}
The residue at the positive-root pole is
\begin{equation}
\operatorname{Res}_{m,+}F
=\frac{\ee^{-\ii\nu_0\alpha}
\ee^{-\pi m\nu_0}}
{2\rho\sqrt{1+\rho^2}},
\end{equation}
while the residue at the negative-root pole is
\begin{equation}
\operatorname{Res}_{m,-}F
=-\frac{\ee^{\ii\nu_0\alpha}
\ee^{-\pi m\nu_0}}
{2\rho\sqrt{1+\rho^2}}.
\end{equation}
Adding the two residues at the same level gives
\begin{align}
\operatorname{Res}_{m,+}F
+\operatorname{Res}_{m,-}F
&=\frac{
\ee^{-\ii\nu_0\alpha}-\ee^{\ii\nu_0\alpha}}
{2\rho\sqrt{1+\rho^2}}
\ee^{-\pi m\nu_0}\\
&=-\frac{\ii\sin(\nu_0\alpha)}
{\rho\sqrt{1+\rho^2}}
\ee^{-\pi m\nu_0}.
\end{align}
Summing over all $m\ge1$ and using
\begin{equation}
\sum_{m=1}^{\infty}\ee^{-\pi m\nu_0}
=\frac1{\ee^{\pi\nu_0}-1}
=\frac1{\ee^{2\pi\Omega/H}-1},
\end{equation}
which gives
\begin{equation}
\sum\operatorname{Res}F
=-\frac{\ii\sin(\nu_0\alpha)}
{\rho\sqrt{1+\rho^2}}
\frac1{\ee^{2\pi\Omega/H}-1}.
\label{app:sumres}
\end{equation}
The contour in the lower half-plane is clockwise, so the integral along the real axis equals $-2\pi\ii$ times the sum of the residues in the lower half-plane:
\begin{equation}
\int_{-\infty}^{\infty}\dd u\,F(u)
=-\frac{2\pi\sin(\nu_0\alpha)}
{\rho\sqrt{1+\rho^2}}
\frac1{\ee^{2\pi\Omega/H}-1}.
\end{equation}
Substituting this into Eq.~\eqref{app:udS-int} gives
\begin{equation}
\RAB^{\dS}(T,\Omega)
=\frac{H}{4\pi}
\frac{\sin(\nu_0\alpha)}
{\rho\sqrt{1+\rho^2}}
\frac1{\ee^{2\pi\Omega/H}-1}.
\end{equation}
Using further
\begin{equation}
\rho=\frac{Hr}{2},
\qquad
\nu_0\alpha
=\frac{2\Omega}{H}
\asinh\frac{Hr}{2},
\end{equation}
we obtain
\begin{equation}
\RAB^{\dS}(T,\Omega)
=\frac1{2\pi r\sqrt{1+(Hr/2)^2}}
\frac{
\sin\!\left[
\dfrac{2\Omega}{H}
\asinh\!\left(\dfrac{Hr}{2}\right)
\right]
}
{\ee^{2\pi\Omega/H}-1}.
\label{app:Rds-final}
\end{equation}
Multiplying and dividing by $\Omega$, this can be written as
\begin{equation}
\RAB^{\dS}
=
\frac{\Omega}{2\pi}
\frac1{\ee^{2\pi\Omega/H}-1}
\frac{
\sin\!\left[
\dfrac{2\Omega}{H}
\asinh\!\left(\dfrac{Hr}{2}\right)
\right]
}
{\Omega r\sqrt{1+(Hr/2)^2}},
\end{equation}
The last factor is precisely $f_H$ defined in Eq.~\eqref{eq:fH-main} of the main text, so this expression is equivalent to Eq.~\eqref{eq:Rds-factor}.

The local limit can also be obtained directly from this final result. Let $u=Hr/2$; then
\begin{equation}
\frac{2\Omega}{H}\asinh u
=\Omega r+O(r^3),
\end{equation}
we have
\begin{equation}
\sin\!\left[
\frac{2\Omega}{H}\asinh u
\right]
=\Omega r+O(r^3).
\end{equation}
Meanwhile,
\begin{equation}
\Omega r\sqrt{1+u^2}
=\Omega r+O(r^3).
\end{equation}
Their ratio tends to unity, and hence
\begin{equation}
\lim_{r\to0}\RAB^{\dS}
=\frac{\Omega}{2\pi}
\frac1{\ee^{2\pi\Omega/H}-1}.
\end{equation}
Thus the single-atom local spectrum follows directly from the zero-separation limit of the two-atom cross spectrum.

\section{Cross spectrum of uniformly accelerated atoms in Minkowski spacetime}\label{app:acc-spectrum}

For the four-dimensional massless scalar field in the Minkowski vacuum, take the Wightman function
\begin{equation}
G^+_{\Mink}(x,x')
=-\frac1{4\pi^2}
\frac1{(t-t'-\ii0^+)^2
-(x-x')^2-(y-y')^2-(z-z')^2}.
\label{app:Mink-W}
\end{equation}
The two worldlines are those of Eq.~\eqref{eq:acc-trajectories}. For one event on worldline $A$ and one on worldline $B$, define
\begin{equation}
T=\frac{\tau+\tau'}2,
\qquad
s=\tau-\tau'.
\end{equation}
The time difference is
\begin{align}
\Delta t
&=a^{-1}
\left[\sinh(a\tau)-\sinh(a\tau')\right]\\
&=\frac{2}{a}
\cosh(aT)\sinh\frac{as}{2},
\label{app:acc-dt}
\end{align}
The coordinate difference along the acceleration direction is
\begin{align}
\Delta x
&=a^{-1}
\left[\cosh(a\tau)-\cosh(a\tau')\right]\\
&=\frac{2}{a}
\sinh(aT)\sinh\frac{as}{2}.
\label{app:acc-dx}
\end{align}
The transverse coordinate differences are
\begin{equation}
\Delta y=-L,
\qquad
\Delta z=0.
\end{equation}
The time and acceleration-direction contributions to the Minkowski interval therefore combine as
\begin{align}
(\Delta t)^2-(\Delta x)^2
&=\frac{4}{a^2}
\sinh^2\frac{as}{2}
\left[\cosh^2(aT)-\sinh^2(aT)\right]\\
&=\frac{4}{a^2}
\sinh^2\frac{as}{2}.
\end{align}
Using $\cosh^2x-\sinh^2x=1$, the midpoint time $T$ cancels, showing that this transversely separated uniformly accelerated configuration is stationary. Including the transverse separation $L$ gives
\begin{equation}
(\Delta t)^2-(\Delta x)^2-(\Delta y)^2
=\frac4{a^2}
\left[
\sinh^2\frac{as}{2}
-\left(\frac{aL}{2}\right)^2
\right].
\end{equation}
Substitution into Eq.~\eqref{app:Mink-W}, with the boundary prescription restored, gives
\begin{equation}
G^+_{AB,\acc}(s)
=-\frac{a^2}{16\pi^2}
\frac1{
\sinh^2\!\left[\dfrac{a(s-\ii0^+)}2\right]
-(aL/2)^2}.
\end{equation}
This is Eq.~\eqref{eq:Gacc-cross} in the main text.

Its Fourier integral is isomorphic to that in \appref{app:ds-contour}. It is sufficient to make the replacements
\begin{equation}
H\to a,
\qquad
r\to L,
\end{equation}
to obtain
\begin{equation}
\RAB^{\acc}(\Omega)
=\frac1{2\pi L\sqrt{1+(aL/2)^2}}
\frac{
\sin\!\left[
\dfrac{2\Omega}{a}
\asinh\!\left(\dfrac{aL}{2}\right)
\right]
}
{\ee^{2\pi\Omega/a}-1}.
\label{app:Racc}
\end{equation}
Multiplying and dividing by $\Omega$ gives
\begin{equation}
\RAB^{\acc}(\Omega)
=\frac{\Omega}{2\pi}
\frac1{\ee^{2\pi\Omega/a}-1}
f_a(\Omega,L),
\end{equation}
where $f_a$ is precisely that defined in Eq.~\eqref{eq:fa-main} of the main text. Taking $L\to0$, the same limit immediately yields
\begin{equation}
\Rloc^{\acc}(\Omega)
=\frac{\Omega}{2\pi}
\frac1{\ee^{2\pi\Omega/a}-1}.
\end{equation}
This is the Unruh excitation spectrum in our Fourier convention. No thermal bath has been assumed here; the Planck factor arises from the periodic pole structure of the uniformly accelerated worldline in complex time.

\section{Minkowski thermal-bath cross spectrum and local kernel}\label{app:thermal-spectrum}

Consider a free massless real scalar field in Minkowski spacetime. Its mode expansion can be written as
\begin{equation}
\phi(t,\vec x)
=\int\frac{\dd^3\vec k}{(2\pi)^{3/2}\sqrt{2k}}
\left[
a_{\vec k}\ee^{-\ii kt+\ii\vec k\cdot\vec x}
+a_{\vec k}^{\dagger}\ee^{\ii kt-\ii\vec k\cdot\vec x}
\right],
\qquad k=|\vec k|.
\label{app:mode-expansion}
\end{equation}
The thermal state satisfies
\begin{equation}
\langle a_{\vec k}^{\dagger}a_{\vec k'}\rangle_\beta
=n_k\delta^{(3)}(\vec k-\vec k'),
\end{equation}
and
\begin{equation}
\langle a_{\vec k}a_{\vec k'}^{\dagger}\rangle_\beta
=(1+n_k)\delta^{(3)}(\vec k-\vec k'),
\end{equation}
where
\begin{equation}
n_k=\frac1{\ee^{\beta k}-1}.
\end{equation}
Take two static atoms separated by $L$, at spatial positions
\begin{equation}
\vec x_A=(0,0,0),
\qquad
\vec x_B=(0,L,0).
\label{app:thermal-static-positions}
\end{equation}
For a static worldline in Minkowski coordinates, $\dd\vec x=0$, so the line element reduces to $\dd\tau^2=\dd t^2$. Choosing the same time origin, we may set
\begin{equation}
t=\tau,
\qquad
t'=\tau',
\qquad
s=\tau-\tau'=t-t'.
\label{app:thermal-proper-time}
\end{equation}
Thus, throughout the paper, $s$ in the Fourier transform denotes the proper-time difference between the detector events associated with the two field operators; here the proper-time difference equals the coordinate-time difference because the atoms are static in the Minkowski frame. Pulling the mode expansion back to these static worldlines and taking the thermal expectation value gives
\begin{equation}
G^+_{AB,\thm}(s,L)
=\int\frac{\dd^3\vec k}{(2\pi)^3 2k}
\left[(1+n_k)\ee^{-\ii ks}
+n_k\ee^{\ii ks}\right]
\ee^{\ii\vec k\cdot(\vec x_A-\vec x_B)}.
\label{app:thermal-w-before-angle}
\end{equation}
Choose the polar axis of momentum-space spherical coordinates along the line joining the atoms. Reversing the orientation of $\vec x_A-\vec x_B$ changes only the overall sign in the phase exponent and leaves the angular integral unchanged, so we may take
\begin{equation}
\vec k\cdot(\vec x_A-\vec x_B)=kL\cos\theta,
\end{equation}
Hence
\begin{align}
\int\dd\Omega_{\vec k}\,
\ee^{\ii kL\cos\theta}
&=2\pi\int_0^\pi\dd\theta\,
\sin\theta\ee^{\ii kL\cos\theta}\\
&=2\pi\int_{-1}^{1}\dd u\,
\ee^{\ii kLu}\\
&=4\pi\frac{\sin(kL)}{kL}.
\label{app:angle-int}
\end{align}
Furthermore, since
\begin{equation}
\dd^3\vec k=k^2\dd k\dd\Omega_{\vec k},
\end{equation}
we have
\begin{equation}
\frac{k^2}{(2\pi)^3 2k}
\times4\pi
=\frac{k}{4\pi^2}.
\end{equation}
Therefore
\begin{equation}
G^+_{AB,\thm}(s,L)
=\int_0^\infty\dd k\,
\frac{k}{4\pi^2}
\frac{\sin(kL)}{kL}
\left[(1+n_k)\ee^{-\ii ks}
+n_k\ee^{\ii ks}\right].
\end{equation}

Now Fourier transform with respect to $s$:
\begin{align}
\RAB^{\thm}(\Omega,L)
&=\int_{-\infty}^{\infty}\dd s\,
\ee^{-\ii\Omega s}G^+_{AB,\thm}(s,L)\\
&=\int_0^\infty\dd k\,
\frac{k}{4\pi^2}
\frac{\sin(kL)}{kL}
\Bigg[
(1+n_k)
\int_{-\infty}^{\infty}\dd s\,
\ee^{-\ii(\Omega+k)s}
\nonumber\\
&\hspace{7em}
+n_k
\int_{-\infty}^{\infty}\dd s\,
\ee^{-\ii(\Omega-k)s}
\Bigg].
\label{app:thermal-fourier-full}
\end{align}
Using the distributional identity
\begin{equation}
\int_{-\infty}^{\infty}\dd s\,
\ee^{-\ii q s}
=2\pi\delta(q),
\end{equation}
we obtain
\begin{equation}
\RAB^{\thm}
=\frac1{2\pi}
\int_0^\infty\dd k\,
k\frac{\sin(kL)}{kL}
\left[(1+n_k)\delta(\Omega+k)
+n_k\delta(\Omega-k)\right].
\end{equation}
Because $k\ge0$ and $\Omega>0$, the root $k=-\Omega$ of $\Omega+k=0$ lies outside the integration range, so the first term vanishes. For the second term, use
\begin{equation}
\int_0^\infty\dd k\,F(k)\delta(k-\Omega)
=F(\Omega)
\end{equation}
which gives
\begin{equation}
\RAB^{\thm}(\Omega,L)
=\frac{\Omega}{2\pi}
n_\Omega
\frac{\sin(\Omega L)}{\Omega L}.
\end{equation}
Substituting
\begin{equation}
n_\Omega=\frac1{\ee^{\beta\Omega}-1},
\end{equation}
yields
\begin{equation}
\RAB^{\thm}(\Omega,L)
=\frac{\Omega}{2\pi}
\frac1{\ee^{\beta\Omega}-1}
\sinc(\Omega L).
\end{equation}
Taking $L\to0$ and using $\sinc0=1$ gives
\begin{equation}
\Rloc^{\thm}(\Omega)
=\frac{\Omega}{2\pi}
\frac1{\ee^{\beta\Omega}-1}.
\end{equation}

The local thermal kernel can be summed directly using the imaginary-time image representation. At a fixed spatial point, the finite-temperature Wightman function is
\begin{equation}
G^+_{\mathrm{loc},\thm}(s)
=
-\frac1{4\pi^2}
\sum_{n=-\infty}^{\infty}
\frac1{(s-\ii n\beta-\ii0^+)^2}.
\label{app:thermal-image-sum}
\end{equation}
Using the complex-analysis identity
\begin{equation}
\sum_{n=-\infty}^{\infty}
\frac1{(z-\ii n\beta)^2}
=
\frac{\pi^2}{\beta^2}
\csch^2\!\left(\frac{\pi z}{\beta}\right),
\label{app:csch-sum}
\end{equation}
and setting $z=s-\ii0^+$ gives
\begin{equation}
G^+_{\mathrm{loc},\thm}(s)
=
-\frac1{4\beta^2}
\csch^2\!\left[
\frac{\pi(s-\ii0^+)}{\beta}
\right].
\label{app:Gth-local-closed}
\end{equation}
When $\beta=2\pi/H$,
\begin{equation}
G^+_{\mathrm{loc},\thm}(s)
=
-\frac{H^2}{16\pi^2}
\csch^2\!\left[
\frac{H(s-\ii0^+)}2
\right].
\end{equation}
On the other hand, the $r\to0$ limit in \appref{app:ds-kernel} and the $L\to0$ limit in \appref{app:acc-spectrum} give exactly the same local kernels for de Sitter spacetime and accelerated Minkowski motion. Thus Eq.~\eqref{eq:triple-local-kernel} in the main text states that the three local Wightman kernels themselves are pointwise identical, not merely that their Fourier transforms happen to yield the same local spectrum.

\section{Asymptotic behavior of the spatial modulation factors}\label{app:spatial-asymptotics}

To derive the short-distance expansion and long-distance asymptotics used in the main text, let
\begin{equation}
x=\Omega L,
\qquad
h=HL.
\end{equation}
For $h\ll1$,
\begin{equation}
\asinh\frac h2
=\frac h2-\frac{h^3}{48}+O(h^5),
\end{equation}
so the phase is
\begin{align}
\frac{2\Omega}{H}\asinh\frac h2
&=\frac{2x}{h}
\left(\frac h2-\frac{h^3}{48}+O(h^5)\right)\\
&=x-\frac{x h^2}{24}+O(h^4).
\end{align}
Hence
\begin{align}
\sin\!\left(x-\frac{x h^2}{24}\right)
&=\sin x-\frac{x h^2}{24}\cos x+O(h^4),\\
\sqrt{1+\frac{h^2}{4}}
&=1+\frac{h^2}{8}+O(h^4).
\end{align}
Therefore
\begin{align}
f_H
&=\frac{
\sin x-\dfrac{x h^2}{24}\cos x+O(h^4)}
{x\left[1+\dfrac{h^2}{8}+O(h^4)\right]}\\
&=\frac{\sin x}{x}
-h^2\left[
\frac18\frac{\sin x}{x}
+\frac1{24}\cos x
\right]
+O(h^4),
\end{align}
which is Eq.~\eqref{eq:small-HL} in the main text.

For $h\gg1$, use
\begin{equation}
\asinh z
=\ln\left(z+\sqrt{z^2+1}\right)
\end{equation}
and set $z=h/2$ to obtain
\begin{equation}
\asinh(h/2)
=\ln h+O(h^{-2}).
\end{equation}
Meanwhile,
\begin{equation}
\sqrt{1+h^2/4}
=\frac h2\left[1+O(h^{-2})\right].
\end{equation}
Therefore
\begin{equation}
f_H(\Omega,L)
=\frac{2}{\Omega H L^2}
\sin\!\left[
\frac{2\Omega}{H}\ln(HL)+O((HL)^{-2})
\right]
\left[1+O((HL)^{-2})\right],
\end{equation}
whose oscillation envelope is $O(L^{-2})$. By contrast,
\begin{equation}
\sinc(\Omega L)=\frac{\sin(\Omega L)}{\Omega L}
\end{equation}
has an $O(L^{-1})$ envelope. Thus the long-distance cross correlations in a Minkowski thermal bath and in the de Sitter/transversely accelerated cases have different power-law decays.

\section{Static de Sitter comparison}\label{app:static-ds}

As a stationary comparison, consider two atoms in the static patch of de Sitter spacetime. The four-dimensional static metric is
\begin{equation}
\dd s^2
=\left(1-H^2R^2\right)\dd t^2
-\frac{\dd R^2}{1-H^2R^2}
-R^2\dd\Omega_2^2,
\qquad 0\le R<H^{-1}.
\end{equation}
Let the two atoms be located at the same static radius $R$, differing only in angular position. Define
\begin{equation}
\kappa
=\frac{\sqrt{1-H^2R^2}}{H}
=\sqrt{H^{-2}-R^2},
\label{app:kappa-static}
\end{equation}
The proper time of a static atom then satisfies $\dd\tau=\sqrt{1-H^2R^2}\,\dd t$, and its local temperature is
\begin{equation}
T_{\rm loc}=\frac1{2\pi\kappa}.
\end{equation}
Let the angular separation of the two points be $\Delta\theta$, and define the chord-length parameter that arises naturally in the static embedding coordinates
\begin{equation}
L_{\rm ch}=2R\sin\frac{\Delta\theta}{2}.
\label{app:Lch}
\end{equation}
It is important to distinguish $L_{\rm ch}$ from the proper arc length along the sphere of fixed $R$ on a static slice. For $0\le\Delta\theta\le\pi$, the corresponding angular arc length is
\begin{equation}
L_{\rm arc}=R\Delta\theta.
\label{app:Larc}
\end{equation}
In general, $L_{\rm ch}\neq L_{\rm arc}$; they agree only to leading order when $\Delta\theta\ll1$. Existing calculations for two static atoms in de Sitter spacetime show that the cross-spectrum function is controlled by $L_{\rm ch}$\cite{HuYu2013,TianWangJingDragan2016,LiuTianWangJing2018}.
Substituting the static trajectories into the conformal-scalar Wightman function of the de Sitter-invariant vacuum gives\cite{HuYu2013,TianWangJingDragan2016,LiuTianWangJing2018}
\begin{equation}
G^+_{\rm loc,stat}(s)
=-\frac1{16\pi^2\kappa^2}
\frac1{\sinh^2[(s-\ii0^+)/(2\kappa)]},
\label{app:static-ds-local}
\end{equation}
and
\begin{equation}
G^+_{AB,\rm stat}(s)
=-\frac1{16\pi^2\kappa^2}
\frac1{
\sinh^2[(s-\ii0^+)/(2\kappa)]
-(L_{\rm ch}/2\kappa)^2}.
\label{app:static-ds-cross}
\end{equation}
where we used
\begin{equation}
\frac{R^2}{\kappa^2}
\sin^2\frac{\Delta\theta}{2}
=\frac{L_{\rm ch}^2}{4\kappa^2}.
\end{equation}
The contour integral of Eq.~\eqref{app:static-ds-cross} is identical to the geometric-type integral in the main text, with the replacements
\begin{equation}
H\longrightarrow\kappa^{-1},
\qquad
r\longrightarrow L_{\rm ch}.
\end{equation}
With the positive-gap excitation Fourier convention used here, this gives
\begin{equation}
\mathcal R_{AB}^{\rm dS,stat}(\Omega)
=\frac{\Omega}{2\pi}
\frac1{\ee^{2\pi\kappa\Omega}-1}
\frac{
\sin[2\kappa\Omega\asinh(L_{\rm ch}/2\kappa)]
}{
\Omega L_{\rm ch}\sqrt{1+(L_{\rm ch}/2\kappa)^2}
}.
\label{app:static-ds-spectrum}
\end{equation}
The proper acceleration of a static de Sitter observer is
\begin{equation}
a_{\rm stat}
=\frac{H^2R}{\sqrt{1-H^2R^2}},
\end{equation}
and satisfies the Deser--Levin-type relation
\begin{equation}
\kappa^{-2}=H^2+a_{\rm stat}^2.
\label{app:kappa-acc-relation}
\end{equation}
Therefore, the Minkowski acceleration matching $a=\kappa^{-1}$ used below matches the total local temperature of the static atom; it does not, at a generic $R$, set the Minkowski acceleration equal to $a_{\rm stat}$. The two become asymptotically equal only near the cosmological horizon.

Now consider two transversely separated uniformly accelerated atoms in the Minkowski vacuum and set
\begin{equation}
a=\kappa^{-1}.
\end{equation}
Their Unruh temperature is then also $(2\pi\kappa)^{-1}$. Further impose the Minkowski transverse proper separation
\begin{equation}
L_{\rm M}=L_{\rm ch}.
\end{equation}
The accelerated spectrum in the main text then becomes
\begin{equation}
\mathcal R_{AB}^{\rm acc}(\Omega; a=\kappa^{-1},L_{\rm M}=L_{\rm ch})
=\mathcal R_{AB}^{\rm dS,stat}(\Omega;\kappa,L_{\rm ch}).
\label{app:static-acc-degeneracy}
\end{equation}
Thus, when both the local-temperature parameter and the spatial parameter in the cross kernel are matched, the two stationary configurations yield cross spectra of exactly the same form. The geometric matching must be distinguished carefully: $L_{\rm M}$ in Minkowski spacetime is a transverse proper distance, whereas $L_{\rm ch}$ in de Sitter spacetime is the chord-length parameter of the static embedding geometry. If instead one matches the proper arc length $L_{\rm arc}$ on the static slice, Eq.~\eqref{app:static-acc-degeneracy} is in general no longer identical term by term. This result concerns only the two-point Wightman cross spectrum and does not exclude the possibility that other observables distinguish de Sitter and Unruh thermality\cite{ZhouChengYu2023}. By contrast, $r(T)$ in the comoving configuration of the main text is the equal-time proper distance on the spatially flat slice and changes with cosmic expansion.

\end{document}